**Oleg Chertov, Dan Tavrov**
National Technical University of Ukraine
"Kyiv Politechnic Institute",
Applied Mathematics Department

# GROUP ANONYMITY: PROBLEMS AND SOLUTIONS

**Existing methods of providing data anonymity preserve individual privacy, but, the task of protecting respondent groups' information in publicly available datasets remains open. Group anonymity lies in hiding (masking) data patterns that cannot be revealed by analyzing individual records. We discuss main corresponding problems, and provide methods for solving each one.**

**Сучасні методи анонімізації даних забезпечують індивідуальну приватність, проте задача захисту інформації про групи респондентів у публічно доступних даних залишається відкритою. Задача групової анонімності – приховати (замаскувати) властивості даних, які не можна визначити на основі аналізу індивідуальних записів. Ми розглядаємо основні пов'язані з цим задачі та пропонуємо методи розв'язку кожної з них.**

**Ключові слова – групова анонімність, вейвлет-перетворення.**

## Introduction

Year to year, the amount of the digital data grows rapidly [1], and so does the possibility to access them. By analyzing primary data, the researchers can get much more information than it is contained in different tables, charts or graphs built using these data even for the most detailed reports. Moreover, they can receive an opportunity to find out different patterns and distribution features that weren't distinguished at the time of the first data publishing. Also, it may be very useful to compare the primary data obtained from different sources.

In this paper, we suppose the data owner releases a depersonalized microfile. That is a file consisting of data records, and each of them contains some attribute (variable) values interesting for the researches but no direct identifiers (such as Social Security Number or full name). Sometimes such microfiles can be publicly available. For instance, sample 5 per cent data on the U.S. Census-2000 [2] can be easily accessed via the Internet. But, in most cases, to gain access to the data, it is required to explain to the data owner the purpose of the planned research (by filling a simple request form). For instance, you need to do this if you want to get an access to 279 million person records collected from 130 censuses of 44 countries provided by IPUMS [3].

Usually, the access is given not to the complete primary data sets but to the data samples. In addition, some attribute's values are masked somehow, or even are absolutely unavailable. These restrictions are necessary to provide the published data anonymity, which is a subject to special legal regulation in different countries (e.g., see the Health Insurance Portability and Accountability Act of 1996 (HIPAA) [4], and the Patient Safety and Quality Improvement Act of 2005 (PSQIA) [5] concerning the health protection data in the USA, or Directive on privacy and electronic communications [6] about electronic commerce in the EU, or the State Statistics Law [7] about providing confidentiality of the primary statistical information in Ukraine).

Anonymity is derived from the Greek word ανωνυμια meaning "without a name" or "namelessness". In a consolidated proposal for terminology [8], anonymity of a subject means that the subject is not identifiable (uniquely characterized) within a set of subjects. Then, data anonymity, as we understand it in this paper, means that the respondents' records cannot be identified within a microfile.

Based on the model of an anonymous message transmission system there can be distinguished two anonymity classes, namely, global and individual [8] (or local [9]) anonymity. Whether the anonymity belongs to a particular class depends on its purpose. It can be achieved for the users corresponding to the message (senders or recipients) globally (for all such users), or for each user individually. In the terms of the task under review in this paper, such classes won't be differentiated [10], and will be considered as the anonymity of individuals or individual data anonymity (property of being unidentifiable within a group).

We have to pay attention to several other important positions connected with the term of anonymity.

When the systems for anonymous communication were studied, the term of the degree of anonymity was introduced [11]. If to apply this term to the field under review in this paper, we can admit that the degree of anonymity provided against an attacker can be viewed as a continuum, ranging from complete anonymity ("absolute privacy") (when the attacker cannot perceive the presence of some respondents in the data sample) to no anonymity ("probably exposed") (when the attacker can identify the personal respondent record).

While discussing the term of anonymity, it is important to note that there exists a set of similar (and sometimes even overlapping) but different information-hiding requirements [12], [13], [8] such as noninterference, privacy, confidentiality, secrecy, unlinkability etc., but they are not the subject to this paper.

The restrictions required by providing the data anonymity of individuals are as important as comprehensive. That's why they are having been successfully studied by the researchers in such fields as statistical disclosure control [14], privacy-preserving data mining [15], distributed privacy, cryptography and adversarial collaboration [16] for many years now.

Nowadays, the data anonymity is achieved by using perturbative and/or non-perturbative methods for masking microfile data [17]. The perturbative methods modify particular microfile values (for example, data swapping [18], microaggregation [19], rounding [17] and others). At the same time, such methods as generalization, top and bottom coding, and local suppression [14] belong to the non-perturbative ones as they let to mask some value combinations without altering them.

By the way, we can determine two main principles for providing individual data anonymity, i.e. randomization and *k*-anonymity. The core of the data randomization [15], [20] lies in adding noise to the data to mask records' attribute values. *K*-anonymity [10], [21] means that every attribute values' combination corresponds to at least *k* respondents in the microfile with the same combination. The situation when it is impossible to find out a specific individual among a respondent group is sometimes called group anonymity [22]. But, in our opinion, this name is not suitable for this situation because it concerns the individual respondent anonymity inside a specific group rather than the respondent group anonymity itself.

Recently, two novel techniques have been proposed to provide data anonymity. The first one is based on the structure transformations of the data matrix (it could possibly be the microfile data). Two methods representing such a technique are singular value decomposition [23] and nonnegative matrix factorization [24]. The second technique implies using Fourier transformation [25] or wavelet transformation (WT) [26], [27].

But, all the discussed principles and techniques intend to provide mainly the data anonymity of individuals. In [28], the term "group anonymity" has been introduced. The aim of group anonymity is to hide or mask data patterns, features, distributions that cannot be revealed by analyzing individual records only. In other words, group anonymity implies providing anonymity of a respondent group with specific features, or even with particular values or value ranges corresponding to the predefined attributes.

We can define two basic techniques for solving the task of providing group anonymity.

The first one subtends to view this task as the task of providing individual anonymity. To do that, we might introduce an equivalence relation (i.e., "be of the same age") on the respondent set. As a result, we obtain disjoint equivalence classes. Then we can apply individual anonymity methods to the corresponding factor set (for instance, we can hide "100 years old people" among the wider group of "elderly people").

The second technique implies masking the true respondent group distribution over a specific attribute (which we will call a parameter attribute in Section 2.1. For example, we can set a task of masking the regional distribution of the military personnel, or a task of masking the distribution of the diseased people over different strata etc. This technique requires developing totally novel specific methods for providing group anonymity. In this paper, we will actually focus on these precise methods.

But, each of the methods discussed in Section 2.1 for providing group anonymity can be accomplished using two different approaches. We decided to call the first one an extremum transition approach. We will call the second one an "Ali Baba's wife" approach.

An extremum transition approach (according to its name) implies transitting records with specific attribute value combinations from the positions of their extreme quantity to the other possible ones.

The name of the second approach goes back to the Middle Eastern and South Asian stories and folk tales collection named "One Thousand and One Nights". In one of the tales, the thieves marked the Ali Baba's house with a symbol to be able to distinguish it when committing their crime. But, Ali Baba's wife saved her husband from inevitable death by marking all the houses in the neighborhood with the same symbol. In terms of our paper, this approach means that we don't need to eliminate extreme quantity. Instead, we can conceal it by adding several other alleged ones.

On the other hand, when providing group anonymity (as well as individual one) we have to guarantee that the data utility is not reduced significantly. To ensure that, we can use WT special features. Using WT, we can split the primary data into an approximation and multilevel details. Then, to protect data, we can redistribute approximation values, and at the same time prevent utility loss by fixing the details (or altering them only proportionally). To illustrate that, let us refer to [29]. In Russia, the responses to 44 different public polls (1994-2001) showed the following result. Actually, the details reflect hidden time series features which are extremely useful for near- and medium-term social processes forecasting.

*The aim* of the current paper is to analyze the main tasks which arise when providing group anonymity. Along with this, it also proposes methods for solving them which are based on the WT approximation values redistributing with simultaneous fixing the WT details.

The paper is structured as follows. In Section 1 we revise the important WT theory necessary for our further explanation. Then, we discuss different group anonymity tasks in Section 2. In the end, we provide the conclusions and the problems prospective to solve.

## 1. Wavelet Theory Basics

In this paper, all the necessary data modifications are accomplished by using WT which is relatively easy-to-use and powerful enough to serve well in our task. That's why we need to revise those wavelet theory facts that are important for the further explanations. For the detailed information refer to [30], [31].

Let us denote by $s = (s_1, s_2, ..., s_m)$ a discrete signal. In addition, let $h = (h_1, h_2, ..., h_n)$ be a high-pass wavelet decomposition filter, whereas $l = (l_1, l_2, ..., l_n)$ will represent a low-pass wavelet decomposition filter.

We can perform one-level wavelet decomposition the following way:

$$a_1 = s *_{\downarrow 2n} l; \quad d_1 = s *_{\downarrow 2n} h. \tag{1}$$

Here, $*_{\downarrow 2n}$ stands for a convolution with a follow-up dyadic downsampling.

In (1), $a_1$ and $d_1$ are arrays of approximation and detail coefficients at level 1 respectively.

We can also apply one-level wavelet decomposition to approximation coefficients at any level $k-1$. As a result, we will receive approximation and detail coefficients at level $k$. In general, to obtain coefficients at any level $k$, we need to carry out the following operations:

$$a_k = ((\underbrace{s *_{\downarrow 2n} l) ... *_{\downarrow 2n} l}_{k \ times}); \tag{2}$$

$$d_k = (((\underbrace{s *_{\downarrow 2n} l) ... *_{\downarrow 2n} l}_{k-1 \ times}) *_{\downarrow 2n} h). \tag{3}$$

Any signal *s* can be presented as a sum of an appropriate approximation and details:

$$s = A_k + \sum_{i=1}^{k} D_i . \qquad (4)$$

In (4), $A_k$ is an approximation at level *k*, and each $D_i$ is a detail at a particular level *i*.

The connection between approximation and details, on the one hand, and corresponding coefficients, on the other hand, is as follows:

$$A_k = ((a_k \underbrace{*_{\uparrow 2n} l) \dots *_{\uparrow 2n}}_{k\ times} l) ; \qquad (5)$$

$$D_k = (((d_k *_{\uparrow 2n} h) \underbrace{*_{\uparrow 2n} l) \dots *_{\uparrow 2n}}_{k-1\ times} l) . \qquad (6)$$

In these formulae, $*_{\uparrow 2n}$ stands for a dyadic upsampling of the left operand and the follow-up convolution with the right one.

But, (5) and (6) aren't the only possible ways of receiving approximations and details. In [31], it has been shown that such operations can be replaced with matrix multiplications. Particularly, we can slightly extend this approach and obtain $A_k$ as follows:

$$A_k = M_{rec} \cdot a_k . \qquad (7)$$

We will call $M_{rec}$ a wavelet reconstruction matrix (WRM). It can be received by consequent multiplications of appropriate matrices for upsampling and convolution described in [31].

The structure of a WRM is very handy for all the methods under review in this paper, that's why it will be heavily used in the next section.

## 2. Group Anonymity Tasks

In fact, there exist different kinds of tasks for group anonymity which can be solved by similar but a bit different methods. Actually, every next problem can be solved by using a slight modification of a method for solving the previous one. But, before we start discussing the simplest of all group anonymity tasks, we have to introduce important terms.

*Table 1.*

**Microfile Data**

|  | $w_1$ | $w_2$ | ... | $w_\eta$ |
|---|---|---|---|---|
| $r_1$ | $z_{11}$ | $z_{12}$ | ... | $z_{1\eta}$ |
| $r_2$ | $z_{21}$ | $z_{22}$ | ... | $z_{2\eta}$ |
| ... | ... | ... | ... | ... |
| $r_\mu$ | $z_{\mu 1}$ | $z_{\mu 2}$ | ... | $z_{\mu\eta}$ |

So, let the given microfile data be presented in a way similar to Table 1. Here, $\mu$ is the overall number of respondents (records), $\eta$ is the overall number of attributes; $w_j$ stands for the $j^{th}$ attribute, $r_i$ stands for the $i^{th}$ record, $z_{ij}$ stands for the $j^{th}$ attribute value corresponding to the $i^{th}$ record.

The core of each group anonymity method is to redistribute specific elements $z_{ij}$. By doing that, we can guarantee that important group features and patterns are masked.

To start redistributing, we need to decide which microfile elements we'll be eager to redistribute. Since redistributing always implies transmitting particular values over particular value ranges, we need to introduce two important value sets.

Let's denote by $S_v$ a subset of a Cartesian product $w_{v_1} \times w_{v_2} \times ... \times w_{v_l}$ of Table 1 columns, where $v_i$, $i = \overline{1, l}$ are integers. The set itself will be called *a vital set*, and each vector from this set will be called *a* *vital value combination*. Respectively, we will call each element of such a vector *a vital value*, and $w_{v_i}$, $i = \overline{1, l}$ will be called *a vital attribute*.

The vital set represents those attributes that will be used for defining records to be transmitted. E.g., if we wanted to change regional distribution of middle-aged men we would need to pick "Age" and "Sex" as vital attributes. In this case, we would receive a vital value combination ("Male"; "Age 45–65").

Let us also denote by $S_p$ a subset of microfile data elements corresponding to the $p^{th}$ attribute, $p \neq v_i \ \forall i = \overline{1, l}$. This set will be called *a parameter set*. All its elements will be called *parameter values*, whereas the $p^{th}$ attribute will be called *a parameter attribute*.

The parameter set represents an attribute that will serve as a value range to redistribute vital values over. With the case of the middle-aged men, the parameter attribute could possibly be "Country region", "Ethnic group", or "Place of work" depending on the problem definition.

In other words, providing group anonymity actually means redistributing vital value combinations over a value range determined by parameter values.

Now, after having defined necessary value sets, we can proceed to discussing different kinds of group anonymity problems and methods for solving each of them.

**2.1. Quantity Problem.** The first (and the simplest) class represents cases when our aim is to hide or mask extreme quantities of records with a particular vital value combination. In some cases, such extremums can reveal restricted information and lead to its unwanted disclosure. Therefore, we need to redistribute records with these vital value combinations such way that these extremums cannot be found out.

To illustrate the method to be explained further, we took 5-Percent Public Use Microdata Sample Files from the U.S. Census Bureau [2] corresponding to the 2000 U.S. Census microfile data on the state of Florida and set a task of protecting the distribution of military personnel over the regions they work in.

A great aid in solving this task is WT. To be able to apply it, we have to construct a signal representing quantities to be changed. For this purpose, we need to calculate the number of microfile records with every pair of a vital value combination and a parameter value possible. Received quantities can be gathered into a signal $q = (q_1, q_2, ..., q_m)$ which will be called *a quantity signal*.

In our example, the only vital attribute is "Military Service", and the only vital value is "1" which stands for "Active Duty".

Since military personnel will be redistributed over the regions they work in, we took "Place of Work PUMA" (where PUMA stands for "Public Use Microdata Area") as a parameter attribute. We also took the parameter values representing each PUMA area corresponding to Florida, i.e. each $10^{th}$ value in the range 12010–12180.

Considering these attributes, we received a following quantity signal: $q$=(669, 794, 9, 11, 852, 9, 4, 280, 31, 118, 6, 13, 1, 24, 7, 14, 18, 135).

Our next aim is to receive a new quantity signal $\tilde{q} = (\tilde{q}_1, \tilde{q}_2, ..., \tilde{q}_m)$ by altering the wavelet approximation of the initial one. It may sound odd but we cannot change the signal's approximation itself, mainly because the wavelet decomposition of a new signal will yield totally different approximation and details. The only possible option to modify signal's approximation is to modify approximation coefficients. As it follows from (5) and (6), the details do not depend on the approximation coefficients, so modifying them doesn't influence the details.

Performing the signal $q$ wavelet decomposition using the first order Daubechies wavelet filter, we get the following approximation coefficients: $a_1$ =(1034.4972, 14.1421, 608.8189, 200.8183, 105.3589, 13.4350, 17.6777, 14.8492, 108.1873). Also, we can obtain a signal's approximation: $A_1$ =(731.5, 731.5, 10, 10, 430.5, 430.5, 142, 142, 74.5, 74.5, 9.5, 9.5, 12.5, 12.5, 10.5, 10.5, 76.5, 76.5), and a signal's detail: $D_1$ =(–62.5, 62.5, –1, 1, 421.5, –421.5, –138, 138, –43.5, 43.5, –3.5, 3.5, –11.5, 11.5, –3.5, 3.5, –58.5, 58.5). Both the approximation and the detail are presented in Fig. 1.

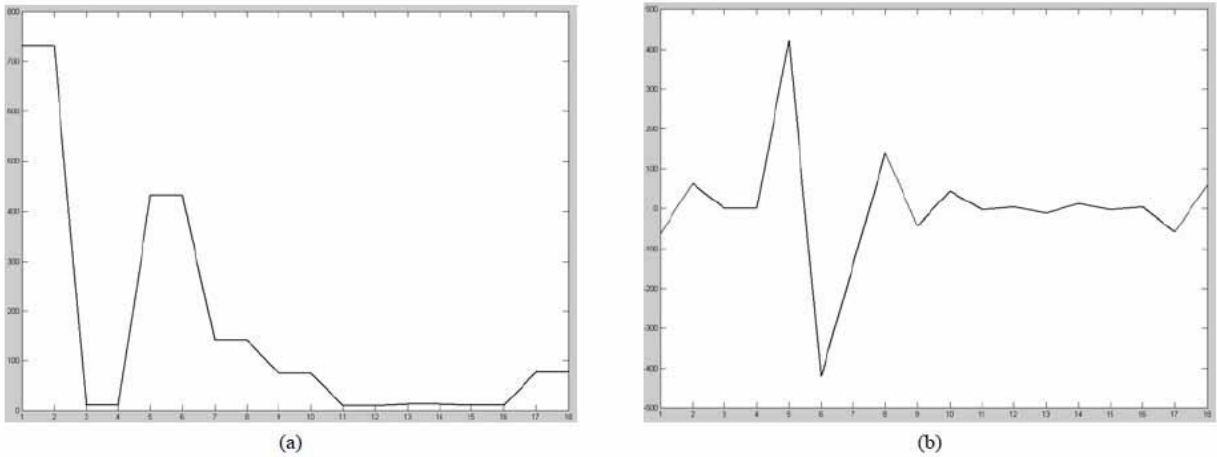

*Fig. 1. Wavelet decomposition of signal q at level 1: a) approximation; b) detail.*

But, how can we possibly define what coefficients to change and how to obtain new approximation with suitable properties? The answer lies in the structure of a corresponding WRM. Taking into consideration (7), we can always determine how to change approximation coefficients to gain needed approximation.

For instance, in our case with the first order Daubechies wavelet filter, we get the WRM presented in Fig. 2.

$$M_{rec} = \begin{pmatrix} 0.7071 & 0 & 0 & 0 & 0 & 0 & 0 & 0 & 0 \\ 0.7071 & 0 & 0 & 0 & 0 & 0 & 0 & 0 & 0 \\ 0 & 0.7071 & 0 & 0 & 0 & 0 & 0 & 0 & 0 \\ 0 & 0.7071 & 0 & 0 & 0 & 0 & 0 & 0 & 0 \\ 0 & 0 & 0.7071 & 0 & 0 & 0 & 0 & 0 & 0 \\ 0 & 0 & 0.7071 & 0 & 0 & 0 & 0 & 0 & 0 \\ 0 & 0 & 0 & 0.7071 & 0 & 0 & 0 & 0 & 0 \\ 0 & 0 & 0 & 0.7071 & 0 & 0 & 0 & 0 & 0 \\ 0 & 0 & 0 & 0 & 0.7071 & 0 & 0 & 0 & 0 \\ 0 & 0 & 0 & 0 & 0.7071 & 0 & 0 & 0 & 0 \\ 0 & 0 & 0 & 0 & 0 & 0.7071 & 0 & 0 & 0 \\ 0 & 0 & 0 & 0 & 0 & 0.7071 & 0 & 0 & 0 \\ 0 & 0 & 0 & 0 & 0 & 0 & 0.7071 & 0 & 0 \\ 0 & 0 & 0 & 0 & 0 & 0 & 0.7071 & 0 & 0 \\ 0 & 0 & 0 & 0 & 0 & 0 & 0 & 0.7071 & 0 \\ 0 & 0 & 0 & 0 & 0 & 0 & 0 & 0.7071 & 0 \\ 0 & 0 & 0 & 0 & 0 & 0 & 0 & 0 & 0.7071 \\ 0 & 0 & 0 & 0 & 0 & 0 & 0 & 0 & 0.7071 \end{pmatrix}$$

*Fig. 2. WRM for reconstructing signals of length 18 using the first order Daubechies wavelet from level 1.*

As we can see, there are extremums in the 2$^{nd}$ and the 5$^{th}$ signal *q* elements. Coming from the WRM structure, we can see that modifying one approximation coefficient results in modifying two neighboring approximation elements. If we take a look at Fig. 1b we will see that the 5$^{th}$ and the 6$^{th}$ detail elements are the reason for the one of the extremums mentioned. All this means we cannot eliminate it completely by changing details only proportionally. That's why the only option left is to apply "Ali Baba's wife" approach. For example, we can enlarge elements of the second half of the signal so that initial maximums are similar to them. Besides, it would be also important to preserve the approximation mean value. All that can be completed if to pick such approximation coefficients: $\hat{a}_1$ =(334.3871, 390.1183, –445.8494,

55.7312, 167.1935, 445.8494, 501.5806, 390.1183, 278.6559). We have to note here that such coefficients are not exlusive and can be substituted with others that fulfil the mentioned requirements.

Using chosen coefficients, we can get a new approximation and a new quantity signal (the signal is obtained by adding the old detail to the new approximation): $\hat{A}_1$ =(236.4474, 236.4474, 275.8553, 275.8553, −315.2632, −315.2632, 39.4079, 39.4079, 118.2237, 118.2237, 315.2632, 315.2632, 354.6711, 354.6711, 275.8553, 275.8553, 197.0395, 197.0395); $\hat{q}$ =(173.9474, 298.9474, 274.8553, 276.8553, 106.2368, −736.7632, −98.5921, 177.4079, 74.7237, 161.7237, 311.7632, 318.7632, 343.1711, 366.1711, 272.3553, 279.3553, 138.5395, 255.5395).

But, as we can see, some signal elements are negative! This is totally unacceptable. To overcome this problem, we need to subtract from every signal's value an arbitrary negative value to make all the elements positive. E.g., in our case we can take −800.

Though, another problem arises. The mean value of the resultant quantity signal differs from the mean value of the original one. This is a great backfire because we can only redistribute military officers not create new ones! To fix this error, we have to multiply our signal by a corresponding coefficient $\sum_{i=1}^{18} q_i / \sum_{i=1}^{18} \hat{q}_i$ =0.1722: $\breve{q}$ =(167.6903, 189.2123, 185.0642, 185.4085, 156.0322, 10.8879, 120.7655, 168.2861, 150.6063, 165.5857, 191.4188, 192.6241, 196.8265, 200.7866, 184.6337, 185.8390, 161.5939, 181.7385).

If to perform the wavelet decomposition of this signal we get the following detail: $\breve{D}_1$ =(−15.6055, 15.6055, −0.2497, 0.2497, 105.2432, −105.2432, −34.4569, 34.4569, −10.8614, 10.8614, −0.8739, 0.8739, −2.8714, 2.8714, −0.8739, 0.8739, −14.6067, 14.6067). As we can see, this detail is actually the initial one multiplied by 4.0050. This means our method actually ensures that the details are changed proportionally.

Finally, we need to round our signal since quantities have to be integers: $\tilde{q}$ =(168, 189, 185, 185, 156, 11, 121, 168, 151, 166, 191, 193, 197, 201, 185, 186, 162, 182).

Alas, we obtain our desired quantity signal with a totally different distribution but with all the wavelet details preserved. Of course, rounding the signal can produce slight changes in the details but these changes do not pose a big threat to preventing utility loss.

Both the initial and the resultant quantity signals are presented in Fig. 3.

After receiving new quantities, the only thing left is to construct a new microfile by changing records' attribute values such way that the number of military personnel in every region is equal to the corresponding element of the quantity signal $\tilde{q}$.

**2.2. Concentration Problem.** Of course, the task from the previous subsection is important but in most cases absolute quantities do not provide the representative information on respondents and thus redistributing them may lead to a serious loss of data utility.

In this subsection we will discuss another class of group anonymity problems that arise mainly when the task is to hide extreme ratios rather than extreme absolute quantities.

The method for solving such problems requires performing one additional signal transformation, though all the other algorithm steps remain the same.

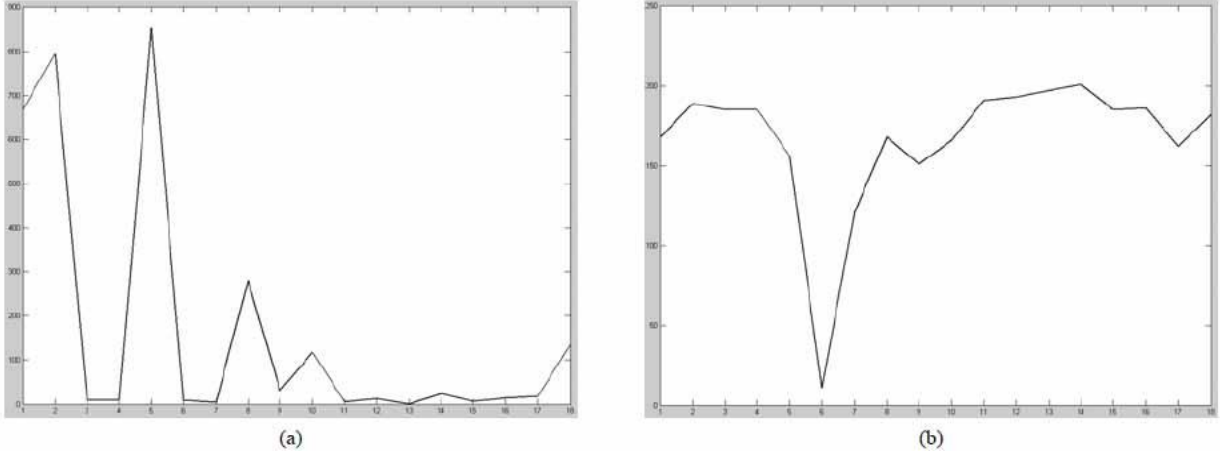

*Fig. 3. Quantity signals for the quantity problem: a) initial; b) resultant.*

To illustrate this method, we will take the same data we took before.

After having constructed a quantity signal (it can be found in the previous subsection), we need to transform it into *a concentration signal* $c = (c_1, c_2, ..., c_m)$. This signal is obtained by dividing every quantity signal's element by the overall number of respondents with the same parameter value and the vital values representing the group of records which comprises the records to be redistributed. In our case, we divided the military personnel quantities by the total number of employed people in each region. We received the following concentration signal: $c$=(0.0799, 0.0738, 0.0009, 0.0010, 0.0331, 0.0009, 0.0006, 0.0055, 0.0008, 0.0113, 0.0006, 0.0014, 0.0001, 0.0021, 0.0006, 0.0006, 0.0006, 0.0029).

The next steps should be carried out as before.

Using the same wavelet filter we used earlier, we get the following approximation coefficients, approximation, and detail: $a_1$ =(0.1087, 0.0014, 0.0240, 0.0043, 0.0085, 0.0015, 0.0016, 0.0009, 0.0025); $A_1$ =(0.0768, 0.0768, 0.0010, 0.0010, 0.0170, 0.0170, 0.0031, 0.0031, 0.0060, 0.0060, 0.0010, 0.0010, 0.0011, 0.0011, 0.0006, 0.0006, 0.0018, 0.0018); $D_1$ =(0.0030, –0.0030, –0.0000, 0.0000, 0.0161, –0.0161, –0.0025, 0.0025, –0.0053, 0.0053, –0.0004, 0.0004, –0.0010, 0.0010, 0.0000, –0.0000, –0.0012, 0.0012).

As we can see, there are extremums in the 1st and the 2nd signal elements. To mask them using "Ali Baba's wife" approach, we can pick such new approximation coefficients: $\hat{a}_1$ =(0.0268, 0.0268, –0.0307, 0.0038, 0.0115, 0.0383, 0.0268, 0.0307, 0.0192). Using (7), we get the following new approximation and concentration signal: $\hat{A}_1$ =(0.0190, 0.0190, 0.0190, 0.0190, –0.0217, –0.0217, 0.0027, 0.0027, 0.0081, 0.0081, 0.0271, 0.0271, 0.0190, 0.0190, 0.0217, 0.0217, 0.0135, 0.0135); $\hat{c}$ =(0.0220, 0.0159, 0.0189, 0.0190, –0.0056, –0.0378, 0.0002, 0.0052, 0.0029, 0.0134, 0.0267, 0.0275, 0.0180, 0.0200, 0.0217, 0.0217, 0.0124, 0.0147).

As in the case with the absolute quantities, we need to make our signal completely positive (by subtracting appropriate negative value, e.g. –0.5).

Now, we need to return to the quantity signal. We can easily accomplish that by multiplying every concentration signal element by the overall number of employed people in each region: $\hat{q}$ =(4371.8096, 5550.8392, 5024.7454, 5636.4255, 12732.8738, 4692.9790, 3459.6336, 25602.6591, 20060.5934, 5366.4042, 4946.7006, 4755.9042, 4549.9511, 5991.3911, 5821.1188, 12584.3353, 15712.6136, 23768.0194).

The next step is to multiply the signal by the coefficient $\sum_{i=1}^{18} q_i / \sum_{i=1}^{18} \hat{q}_i$ =0.0176 in order not to change its mean value: $\bar{q}$ =(76.7371, 97.4322, 88.1979, 98.9345, 223.4963, 82.3745, 60.7259, 449.3959, 352.1176, 94.1949, 86.8280, 83.4790, 79.8639, 105.1651, 102.1764, 220.8891, 275.7988, 417.1930).

We can decompose this signal and get the following detail: $\breve{D}_1$ =(0.0030, –0.0030, –0.0000, 0.0000, 0.0161, –0.0161, –0.0025, 0.0025, –0.0053, 0.0053, –0.0004, 0.0004, –0.0010, 0.0010, 0.0000, –0.0000, –0.0012, 0.0012).

As we see, in this case the resultant detail is totally equal to the initial one.

Of course, we need not to forget to round the result: $\tilde{q}$ =(77, 97, 88, 99, 223, 82, 61, 449, 352, 94, 87, 83, 80, 105, 102, 221, 276, 417).

Both initial and resultant quantity signals are presented in Fig. 4. It is important to note that though the concentration signals were altered mostly the same way the quantity signals from the previous subsection were, the resultant quantities in this example have a completely different distribution. Even applying "Ali Baba's wife" approach to the concentration signals we actually got the extremum transition approach for the corresponding quantity signals. This means modifying concentrations can yield a totally different outcome.

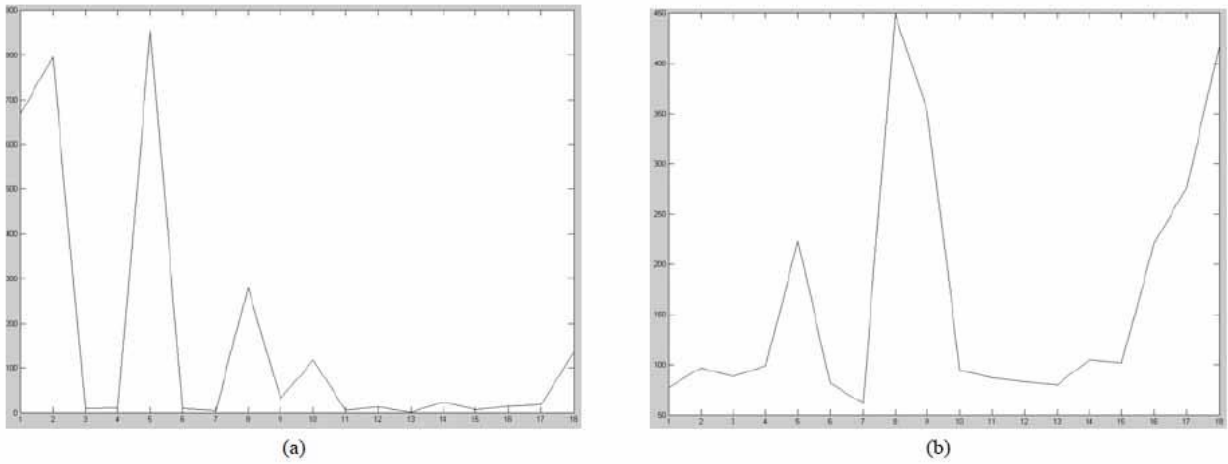

*Fig. 4. Quantity signals for the concentration problem: a) initial; b) resultant.*

**2.3. Concentration Difference Problem.** Though the previous method is absolutely acceptable for solving the most frequent and important practical problems, there is another class of group anonymity tasks which aim mainly at protecting comparative distributions. For example, if there are much more young males than females in a particular region it can possibly be a hint to finding out a location of a military cantonment.

The method we propose to use for solving tasks of this kind is very similar to the two described before. It requires following one additional step but it is free of a "negative elements" trouble.

We will define two different vital value sets. The first one will be called *a main vital set*, and the other one will be called *a subordinate vital set*. Each vector from the main vital set will be called *a main vital value combination*. Respectively, each element of this vector will be called *a main vital value*.

By analogue, each vector from the subordinate vital set will be called *a subordinate vital value combination*, whereas each element of such a vector will be called *a subordinate vital value*.

As in the previous subsections, we need to construct main and subordinate quantity signals taking into account main and subordinate vital value combinations respectively.

To illustrate this new method, we took "Sex" and "Age" as vital attributes. In particular, we took a made-up value "Young age" (which comprises actual "Age" values from "18" to "25") as both main and subordinate vital value; at the same time, we took "Sex" value "1" (standing for "Male") as a main vital value, and value "2" (standing for "Female") as a subordinate vital value.

We received the following quantity signals ($q^1$ stands for the main quantity signal, and $q^2$ stands for the subordinate one): $q^1$=(885, 931, 863, 996, 2014, 683, 435, 3212, 3037, 584, 712, 607, 690, 678, 689,

1458, 1865, 2947); $q^2$ =(591, 713, 876, 982, 1798, 629, 373, 3110, 2725, 579, 584, 452, 438, 580, 578, 1246, 1806, 2821).

According to the previous subsection, we need to construct concentration signals. We decided to divide received quantities by the overall number of people in a particular region, yielding the following signals: $c^1$=(0.1057, 0.0865, 0.0891, 0.0917, 0.0782, 0.0673, 0.0629, 0.0634, 0.0761, 0.0559, 0.0758, 0.0673, 0.0786, 0.0588, 0.0617, 0.0604, 0.0608, 0.0638); $c^2$=(0.0706, 0.0663, 0.0905, 0.0904, 0.0698, 0.0620, 0.0539, 0.0614, 0.0683, 0.0554, 0.0622, 0.0501, 0.0499, 0.0503, 0.0518, 0.0516, 0.0589, 0.0611).

Now we came close to the step which is new to our previous algorithm. We need to build up *a concentration difference signal* as follows: $\delta = (\delta_1, \delta_2, ..., \delta_m) \equiv (c_1^1 - c_1^2, c_2^1 - c_2^2, ..., c_m^1 - c_m^2)$. In our example, we got the following signal: $\delta$ =(0.0351, 0.0203, –0.0013, 0.0013, 0.0084, 0.0053, 0.0090, 0.0020, 0.0078, 0.0005, 0.0136, 0.0172, 0.0287, 0.0085, 0.0099, 0.0088, 0.0019, 0.0027).

Since our target is to redistribute these precise concentration differences, we need to treat this signal the same way we've done before. At first, we need to get the approximation coefficients, approximation, and detail: $a_1$ =(0.0392, –0.0000, 0.0097, 0.0078, 0.0059, 0.0218, 0.0263, 0.0132, 0.0033); $A_1$ =(0.0277, 0.0277, –0.0000, –0.0000, 0.0069, 0.0069, 0.0055, 0.0055, 0.0041, 0.0041, 0.0154, 0.0154, 0.0186, 0.0186, 0.0094, 0.0094, 0.0023, 0.0023); $D_1$ =(0.0074, –0.0074, –0.0013, 0.0013, 0.0015, –0.0015, 0.0035, –0.0035, 0.0037, –0.0037, –0.0018, 0.0018, 0.0101, –0.0101, 0.0006, –0.0006, –0.0004, 0.0004).

As we can see, there are extremums in the 1st and the 13th signal elements. To mask them using "Ali Baba's wife" approach, we can pick such new approximation coefficients: $\hat{a}_1$ =(0, 0.0212, 0.0141, 0.0141, 0.0141, 0.0212, –0.0212, 0.0282, 0.0353). Using (7), we can get the following new approximation and concentration difference signal: $\hat{A}_1$ =(0, 0, 0.0150, 0.0150, 0.0100, 0.0100, 0.0100, 0.0100, 0.0100, 0.0100, 0.0150, 0.0150, –0.0150, –0.0150, 0.0200, 0.0200, 0.0250, 0.0250); $\hat{\delta}$ =(0.0074, –0.0074, 0.0137, 0.0163, 0.0115, 0.0084, 0.0135, 0.0065, 0.0137, 0.0063, 0.0132, 0.0168, –0.0049, –0.0251, 0.0205, 0.0194, 0.0246, 0.0254).

If we perform a wavelet decomposition of the resultant concentration difference signal, we will receive the same detail as we received before.

Since the differences can be negative, we do not need to make our signal completely positive.

The next step is to retrieve new concentrations using the differences above. We can always accomplish that by solving a corresponding linear equation system with $2m$ unknowns and $m$ equations (the equations have to ensure that the differences between new concentrations will be equal to the concentration difference signal elements).

For instance, we can get the following signals: $\hat{c}^1$=(0.0780, 0.0865, 0.1041, 0.1067, 0.0813, 0.0704, 0.0674, 0.0679, 0.0820, 0.0617, 0.0754, 0.0669, 0.0786, 0.0588, 0.0723, 0.0710, 0.0834, 0.0865); $\hat{c}^2$=(0.0706, 0.0940, 0.0905, 0.0904, 0.0698, 0.0620, 0.0539, 0.0614, 0.0683, 0.0554, 0.0622, 0.0501, 0.0834, 0.0839, 0.0518, 0.0516, 0.0589, 0.0611).

Now, we need to return to the quantity signals. We can easily accomplish that task by multiplying each concentration signal element by the overall number of people living in each region: $\hat{q}^1$=(653.1524, 931, 1008.2600, 1158.9168, 2094.6127, 714.7812, 466.0843, 3439.7838, 3269.7099, 644.9776, 707.9116, 603.0753, 690, 678, 807.2578, 1713.6776, 2559.0257, 3992.0670); $\hat{q}^2$=(591, 1010.8445, 876, 982, 1798, 629, 373, 3110, 2725, 579, 584, 452, 732.8917, 966.8440, 578, 1246, 1806, 2821).

The next step is to multiply the signals by the coefficients $\sum_{i=1}^{18} q_i^1 / \sum_{i=1}^{18} \hat{q}_i^1 = 0.8911$ and $\sum_{i=1}^{18} q_i^2 / \sum_{i=1}^{18} \hat{q}_i^2 = 0.9552$ in order not to change their mean values, and to round the results: $\tilde{q}^1$=(582, 830, 898,

1033, 1866, 637, 415, 3065, 2914, 575, 631, 537, 615, 604, 719, 1527, 2280, 3557); $\tilde{q}^2$ =(565, 966, 837, 938, 1717, 601, 356, 2971, 2603, 553, 558, 432, 700, 924, 552, 1190, 1725, 2695).

Both initial and resultant quantity signals are presented in Fig. 5 and Fig. 6. It is important to note that the distribution of the young males hasn't changed significantly. Though, due to the females' distribution we gain the desired outcome.

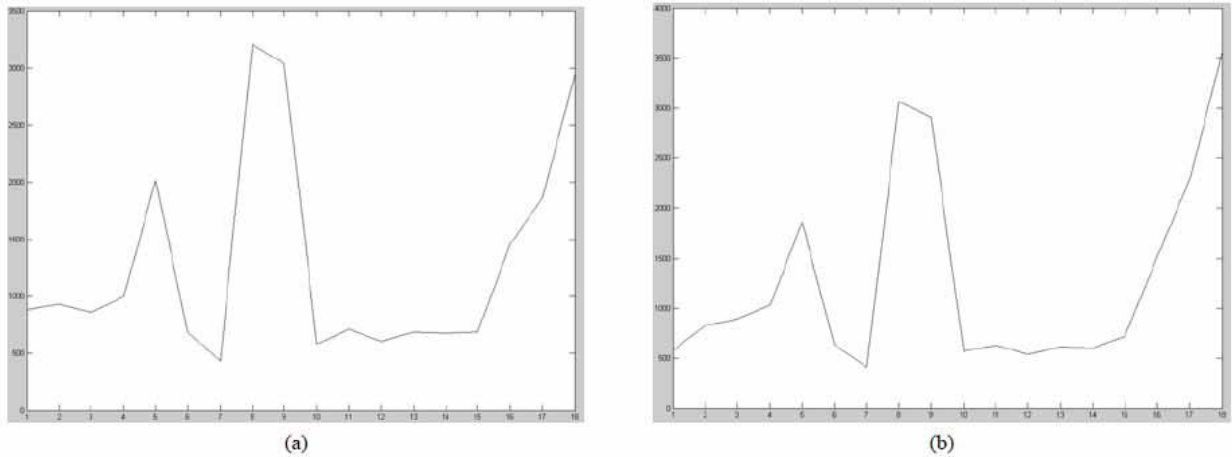

*Fig.5. Quantity signals for the concentration difference problem (males): a) initial; b) resultant.*

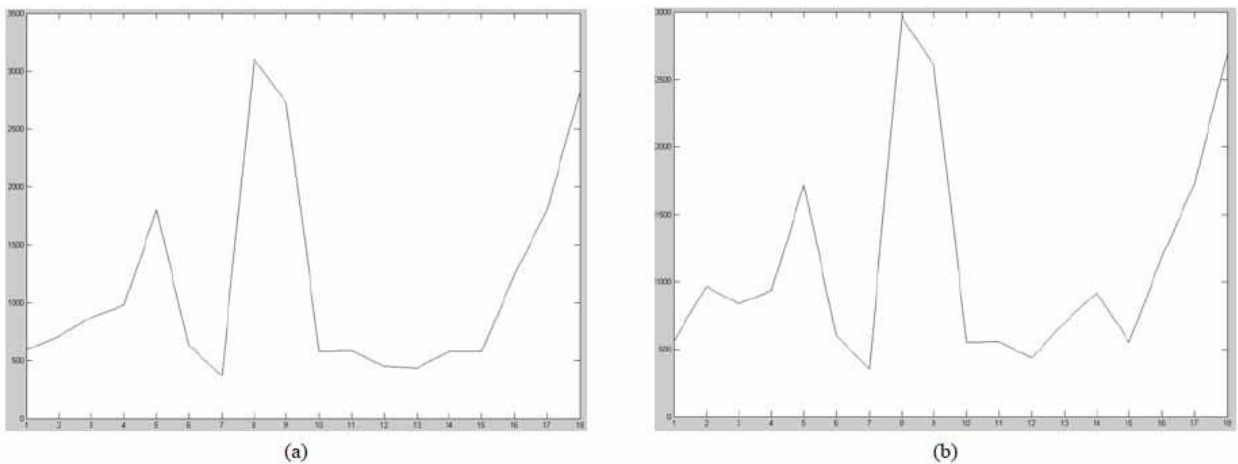

*Fig. 6. Quantity signals for the concentration difference problem (females): a) initial; b) resultant.*

## Conclusions and Future Research

Anonymity is a methodological requirement which lies in the fact that a respondent isn't supposed to be identified (in order to prevent possible usage of the received data against his or her will). In this paper, we found out that the widely used term of data anonymity in fact is used only as a synonym to the anonymity of individuals, i.e. it means that the respondents' records are not identifiable within a dataset. But, in this paper we set a totally different task of providing group anonymity as anonymity of a respondent group. In some cases, this task can be viewed as an individual anonymity task by constructing specific respondent groups and follow-up anonymizing them as separate entities. But, in most other cases the group anonymity task supplies special problems, solving which requires applying novel methods.

In the paper, three main problem kinds immanent to the task under review have been analyzed. They are as follows: providing group anonymity for the absolute quantities, for the ratios and for the differences between the ratios. For each problem kind, detailed examples with different approaches to applying a proposed method have been provided.

The group anonymity problem seems to have a simple solution only at first sight. It seems easy to arbitrarily change the respondent group's distribution over the parameter values (the territory, the age, the ethnic groups etc.). The main question is how much utility will remain in the data after such transformations. To provide balance between protecting privacy and preserving data utility, we proposed a method which is based on using WT. Using it, we can redistribute the WT approximation (to provide anonymity) and fix the WT details or alter them proportionally (to preserve the data utility).

The future researches can lie in introducing the group anonymity measure and in developing optimal algorithms for constructing a microfile with the transformed distribution.

*1. As the economy contracts, the digital universe expands [Електронний ресурс] / J. Gantz, D. Reinsel // IDC multimedia white paper, 2009. — Режим доступу: http://www.emc.com/collateral/demos/microsites/idc-digitaluniverse/iview.htm. 2. U.S. Census 2000. 5-Percent Public Use Microdata Sample Files [Електронний ресурс] / U.S. Census Bureau. — Режим доступу: http://www.census.gov/Press-Release/www/2003/PUMS5.html. 3. Minnesota Population Center. Integrated Public Use Microdata Series International [Електронний ресурс] / IPUMS. — Режим доступу: https://international.ipums.org/international/. 4. Health Insurance Portability and Accountability Act of 1996 (HIPAA): Aug. 21, 1996 / 104[th] Congress. — Public Law 104-191. — Режим доступу: http://www.hipaa.org/. 5. Patient Safety and Quality Improvement Act of 2005 (PSQIA) / Federal Register. — 2001 — 73(266). 6. Directive 2002/58/EC of the European Parliament and of the Council of 12 July 2002: Jul. 31, 2002 / Official Journal of the European Communities. — 2002 — L 201. 7. Закон України "Про державну статистику": станом на 5 бер. 2009. — Режим доступу: http://zakon1.rada.gov.ua/cgi-bin/laws/main.cgi?nreg=2614-12\&p=1265575855780241. 8. A Terminology for Talking about Privacy by Data Minimization: Anonymity, Unlinkability, Undetectability, Unobservability, Pseudonymity, and Identity Management, Version v0.32 [Електронний ресурс] / A. Pfitzmann, M. Hansen. — 2009. — Режим доступу: http://dud.inf.tu-dresden.de/Anon\_Terminology.shtml. 9. Toth G. Measuring Anonymity Revisited / G. Toth, Z. Hornak, F. Vajda // Proceedings of the 9th Nordic Workshop on Secure IT Systems. — Espoo, 2004. — P. 85-90. 10. Sweeney L. Protecting Privacy when Disclosing Information: k-Anonymity and Its Enforcement through Generalization and Suppression / L. Sweeney, P. Samarati // IEEE Symposium on Research in Security and Privacy. — IEEE Computer Society Press, 1998. — P. 86-99. 11. Reiter M. K. Crowds: Anonymity for Web Transactions / M. K. Reiter, A. D. Rubin // ACM Transactions on Information and System Security. — ACM Press, 1998. — 1(1). — P. 66-92. 12. Halpern J. Y. Anonymity and Information Hiding in Multiagent Systems / J. Y. Halpern, K. R. O'Neill // The 16th IEEE Computer Security Foundations Workshop. — IEEE Computer Society Press, 2003. — P. 75-88. 13. Marx G. T. What's in a Name? Some Reflections on the Sociology of Anonymity / G. T. Marx // The Information Society. — Taylor & Francis, 1999. — 15(2). — P. 99-112. 14. Domingo-Ferrer J. A Survey of Inference Control Methods for Privacy-Preserving Data Mining / J. Domingo-Ferrer // Privacy-Preserving Data Mining: Models and Algorithms. — Springer, 2008. — P. 53-80. 15. Agrawal R. Privacy-Preserving Data Mining / R. Agrawal, R. Srikant // ACM SIGMOD International Conference on Management of Data. — ACM Press, 2000. — P. 439-450. 16. Lindell Y. Privacy Preserving Data Mining / Y. Lindell, B. Pinkas // Advances in Cryptology Crypto 2000. — Berlin: Springer, 2000. — Vol. 1880. — P. 36-53. 17. Willenborg L. Elements of Statistical Disclosure Control / L. Willenborg, T. DeWaal. — New York: Springer-Verlag, 2001. — 261 p. 18. Fienberg S. Data Swapping: Variations on a Theme by Dalenius and Reiss / S. Fienberg, J. McIntyre // Journal of Official Statistics. — Stockholm: Almqvist & Wiksell International, 2005. — Vol. 21(2). — P. 309-324. 19. Domingo-Ferrer J. Practical Data-oriented Microaggregation for Statistical Disclosure Control / J. Domingo-Ferrer, J. M. Mateo-Sanz // IEEE Transactions on Knowledge and Data Engineering. — IEEE Computer Society Press, 2002. — 14(1). — P. 189-201. 20. Evfimievski A. Randomization in Privacy Preserving Data Mining / A. Evfimievski // ACM SIGKDD Explorations Newsletter. — ACM Press, 2002. — 4(2). — P. 43-48. 21. Sweeney L. k-anonymity: a Model for Protecting Privacy / L. Sweeney // International Journal on Uncertainty, Fuzziness and Knowledge-based Systems. — World Scientific, 2002. — 10(5). — P. 557-570. 22. Bhargava M. Probabilistic Anonymity /*

*M. Bhargava, C. Palamidessi // CONCUR 2005 – Concurrency Theory. — Berlin/Heidelberg : Springer, 2005. — Vol. 3653. — P. 171-185. 23. Xu S. Singular Value Decomposition Based Data Distortion Strategy for Privacy Protection / S. Xu, J. Zhang, D. Han, J. Wang // Knowledge and Information Systems. — Springer, 2006. — 10(3). — P. 383-397. 24. Wang J. NNMF-based Factorization Techniques for High-Accuracy Privacy Protection on Non-Negative-Valued Datasets / J. Wang, W.J. Zhong, J. Zhang // The 6th IEEE Conference on Data Mining, International Workshop on Privacy Aspects of Data Mining. — IEEE Computer Society Press, 2006. — P. 513-517. 25. Mukherjee S. A Privacy Preserving Technique for Euclidean Distance-based Mining Algorithms Using Fourier-related Transforms / S. Mukherjee, Z. Chen, A. Gangopadhyay // The VLDB Journal. — Springer-Verlag, 2006. — 15(4). — P. 293-315. 26. Bapna S. A Wavelet-based Approach to Preserve Privacy for Classification Mining / S. Bapna, A. Gangopadhyay // Decision Sciences Journal. — Wiley-Blackwell, 2006. — 37(4). — P. 623-642. 27. Liu L. Wavelet-based Data Perturbation for Simultaneous Privacy-Preserving and Statistics-Preserving / L. Liu, J. Wang, J. Zhang // 2008 IEEE International Conference on Data Mining Workshops. — IEEE Computer Society Press, 2008. — P. 27-35. 28. Chertov O. Statistical Disclosure Control Methods for Microdata / O. Chertov, A. Pilipyuk // International Symposium on Computing, Communication and Control. — Singapore: IACSIT, 2009. — P. 338-342. 29. Давыдов А. А. Вейвлет-анализ социальных процессов / А. А. Давыдов // Социологические исследования. — 2003. — №11. — С. 89-101. 30. Mallat S. A Wavelet Tour of Signal Processing / S. Mallat. — New York: Academic Press, 1999. — 620 p. 31. Strang G. Wavelet and Filter Banks / G. Strang, T. Nguyen. — Wellesley: Wellesley-Cambridge Press, 1997. — 520 p.*